\documentclass[a4paper,UKenglish,cleveref, autoref, thm-restate]{lipics-v2021}
\nolinenumbers

\usepackage{listings}
\usepackage{tcolorbox}
\usepackage{xcolor}
\usepackage{tikz}
\usetikzlibrary{arrows.meta,positioning}
\definecolor{delim}{RGB}{20,105,176}
\definecolor{numb}{RGB}{106, 109, 32}
\definecolor{string}{rgb}{0.64,0.08,0.08}
\usepackage{booktabs}
\usepackage{array}
\usepackage{ragged2e}
\newcolumntype{P}[1]{>{\RaggedRight\arraybackslash}p{#1}}

\lstdefinelanguage{json}{
    showspaces=false,
    showtabs=false,
    breaklines=true,
    postbreak=\raisebox{0ex}[0ex][0ex]{\ensuremath{\color{gray}\hookrightarrow\space}},
    breakatwhitespace=true,
    basicstyle=\ttfamily\small,
    upquote=true,
    morestring=[b]",
    stringstyle=\color{string},
    literate=
     *{0}{{{\color{numb}0}}}{1}
      {1}{{{\color{numb}1}}}{1}
      {2}{{{\color{numb}2}}}{1}
      {3}{{{\color{numb}3}}}{1}
      {4}{{{\color{numb}4}}}{1}
      {5}{{{\color{numb}5}}}{1}
      {6}{{{\color{numb}6}}}{1}
      {7}{{{\color{numb}7}}}{1}
      {8}{{{\color{numb}8}}}{1}
      {9}{{{\color{numb}9}}}{1}
      {\{}{{{\color{delim}{\{}}}}{1}
      {\}}{{{\color{delim}{\}}}}}{1}
      {[}{{{\color{delim}{[}}}}{1}
      {]}{{{\color{delim}{]}}}}{1},
}

\newif\iflongversion

\NewDocumentEnvironment{longversion}{ +b }{\iflongversion#1\fi}{}

\EventEditors{}
\EventLongTitle{20th Int.\ Symposium on Empirical Software Engineering and Measurement (ESEM 2026) -- Registered Reports}
\EventShortTitle{ESEM 2026}
\EventAcronym{ESEM}
\EventYear{2026}
\EventDate{October 4--9, 2026}
\EventLocation{Munich, Germany}
\SeriesVolume{}
\ArticleNo{}

\title{The Prompt Triangle: A Registered Report on Prompts as Hybrid Artifacts}

\titlerunning{The Prompt Triangle}

\author{Shalini Chakraborty}{University of Bayreuth, Germany}{shalini.chakraborty@uni-bayreuth.de}{0000-0002-9466-3766}{}

\author{Jan-Philipp Steghöfer}{XITASO GmbH IT and Software Solutions, Augsburg, Germany}{jan-philipp.steghoefer@xitaso.com}{0000-0003-1694-0972}{}

\authorrunning{Chakraborty and Steghöfer}
\Copyright{Shalini Chakraborty and Jan-Philipp Steghöfer}



\begin{document}

\maketitle

\begin{abstract}
AI-based coding assistants are transforming software development by shifting effort from writing code to crafting prompts that guide code generation. Despite their growing importance, little empirical evidence exists on how prompts function as software engineering artifacts. Building on prior work framing prompts as mixed artifacts combining requirement intent and solution guidance, we build on the Prompt Triangle, a conceptual model describing prompts along three components: Functionality and Quality (requirements), General Solutions (architectural guidance), and Specific Solutions (implementation constraints).
This registered report presents the Stage 1 protocol for a confirmatory study investigating how prompts evolve and how their alignment with requirements engineering (RE) activities predicts development outcomes. We preregister four hypotheses examining prompt evolution, developer characteristics, RE activity alignment, and temporal patterns. We employ a controlled experiment (n=30), community uploads, and mined data using dual-coding to test whether prompting patterns predict development success.
\end{abstract}

\section{Introduction}

Recent advances in AI-assisted software development are reshaping software development practices. Developers interact with AI-based coding assistants through natural language prompts that describe intended functionality, constraints, and design preferences. This interaction shifts development effort from code authoring to \emph{prompt construction}, positioning prompts as central rather than ephemeral artifacts.

Prompts are structured artifacts that encode multiple layers of intent~\cite{meske2025vibecodingreconfigurationintent}, including requirements, architectural decisions, and implementation constraints. Recent work has begun to conceptualise prompts as hybrid artifacts that combine elements of requirements engineering and solution design~\cite{chakraborty2026prompts}. However, empirical understanding of how prompts evolve during development, and how their structure influences outcomes, remains limited.

This gap is particularly relevant for empirical software engineering. While prior studies have evaluated the productivity and correctness of AI-generated code~\cite{barke2023grounded}, fewer have examined the \emph{process} through which developers iteratively refine prompts. In traditional software engineering, artifacts such as requirements specifications and design documents evolve systematically and are known to influence final system quality. Whether similar principles apply to prompts remains an open question.

To address this gap, we build on the \textbf{Prompt Triangle}~\cite{chakraborty2026prompts,shalini2026exploring}, a conceptual model introduced in our prior work that characterises prompts along three dimensions: (1) \emph{Functionality and Quality} (capturing requirements), (2) \emph{General Solutions} (capturing architectural or technological guidance), and (3) \emph{Specific Solutions} (capturing implementation-level constraints). This model provides a structured lens for analysing prompt content and evolution.

In this registered report (Stage~1), we present a primarily confirmatory study that tests preregistered hypotheses about prompt evolution and its relationship to development outcomes. Our four hypotheses address: (i) whether prompt content shifts toward specific solutions across iterations, (ii) whether developer characteristics predict prompting strategies, (iii) whether alignment between prompt content and RE activities predicts code quality, and (iv) whether temporal alignment patterns predict outcomes beyond overall alignment. The controlled experiment simultaneously serves as further construct validation of the Prompt Triangle: if the dimensions cannot be reliably coded ($\kappa < 0.75$) or yield degenerate distributions, this constitutes disconfirming evidence.
%
%
By treating prompts as first-class artifacts, this work establishes a foundation for systematic empirical investigation of AI-assisted software development.

\section{Background and Related Work}

Prompt engineering, i.e., crafting prompts to guide LLMs toward desired outputs~\cite{reynolds2021prompt,chen2023unleashing}, has become central to AI-assisted development. However, prompts often lack the structure of formal requirements~\cite{barke2023grounded}. Developers engage in iterative, conversational refinement~\cite{barke2023grounded}, facing challenges in correctness, trust~\cite{vaithilingam2022expectation}, and translating requirements into appropriate prompts~\cite{VillamizarFKVM25}. This raises a key question: \textbf{How can RE rigor be infused into prompting practices?} Viewing prompts as evolving requirement artifacts bridges informal intentions and structured specification~\cite{vogelsang2024prompting}.

Recent work explores how LLMs reshape RE activities such as elicitation, validation, and traceability~\cite{huang2025prompt}. Emerging research conceptualizes prompts as hybrid artifacts combining requirements and solutions. Our prior work~\cite{chakraborty2026prompts} proposes the Prompt Triangle capturing functionality, general solutions, and specific constraints, yet empirical evidence on component evolution and impact remains limited.

\smallskip
\noindent
\textbf{Research Gap.} Existing studies examine prompting patterns~\cite{dicuffa2025exploringpromptpatternsaiassisted}, quality defects~\cite{siddiq2024fault}, output effectiveness~\cite{della2025prompt}, and RE task applications~\cite{huang2025prompt}, but none systematically model prompts as artifacts encoding requirements \textbf{and} solution decisions. This challenges the SE principle of separating requirements from solutions. We address this through the Prompt Triangle~\cite{chakraborty2026prompts,shalini2026exploring}, a structured framework for decomposing prompts into requirement-oriented (\emph{Functionality and Quality}) and solution-oriented (\emph{General} and \emph{Specific Solutions}) components, enabling empirical analysis of their evolution and impact on development outcomes.

\begin{longversion}
Prompt engineering refers to the practice of carefully crafting and refining prompts to guide large language models (LLMs) toward producing desired outputs~\cite{reynolds2021prompt,chen2023unleashing}. In the context of software development, prompt engineering is increasingly seen as a crucial skill for effectively leveraging chat-based coding assistants. However, while prompts may contain elements of system requirements or design intent, they are often ad hoc and lack the structure of formal requirements~\cite{barke2023grounded}. Barke et al.~\cite{barke2023grounded} show that developers engage in iterative, conversational interactions with code-generating models, shifting the unit of work from code edits to prompt refinement. Other studies highlight both productivity gains and challenges related to correctness, trust, and over-reliance on generated code~\cite{vaithilingam2022expectation}. Many developers struggle to provide the right level of detail and to translate requirements into prompts consistently~\cite{VillamizarFKVM25}. This raises an important question for RE: \textbf{How can the rigor of RE be infused into prompting practices?} Viewing prompts as lightweight but evolving requirement artifacts offers a way to bridge the gap between informal developer intentions and structured requirements specification~\cite{vogelsang2024prompting}. 


The \emph{intersection of requirements engineering (RE) and LLMs} is gaining increasing attention. Recent literature reviews argue that LLMs reshape traditional RE activities such as elicitation, validation, and traceability, while introducing new challenges related to ambiguity and control~\cite{huang2025prompt}. At the same time, RE emphasizes the importance of structured artifacts and the separation of concerns between requirements and solutions.


Emerging work conceptualizes \emph{prompts as hybrid artifacts} that combine requirements and solution information. Our prior work~\cite{chakraborty2026prompts} proposes the Prompt Triangle as a lightweight model capturing functionality, general solution strategies, and specific implementation constraints. However, empirical evidence on how these components evolve and influence development outcomes remains limited.

\smallskip
\noindent
\textbf{Research Gap.} Although prior research has examined prompt engineering in AI-assisted software development, important gaps remain in understanding prompts as structured software engineering artifacts. Existing studies primarily focus on prompting patterns, quality defects, or output effectiveness. Di Cuffa et al.~\cite{dicuffa2025exploringpromptpatternsaiassisted}, e.g., identify recurring linguistic prompt patterns, Siddiq et al.~\cite{siddiq2024fault} analyse ambiguity and context deficiencies in developer prompts, and Della Porta et al.~\cite{della2025prompt} evaluate the effect of prompting strategies on code quality. Huang et al.~\cite{huang2025prompt} explore prompt engineering for requirements engineering tasks, but do not distinguish requirement and solution information within prompts. As a result, no existing work systematically models prompts as artifacts that simultaneously encode requirements and solution decisions. This challenges the foundational SE principle of separating requirements from solutions, establishing that prompt structure and evolution directly impact code quality and must be managed accordingly. Our work addresses this gap through the Prompt Triangle, which provides the first structured framework for decomposing prompts into requirement-oriented (\emph{Functionality and Quality}) and solution-oriented (\emph{General} and \emph{Specific Solutions}) components~\cite{shalini2026exploring}, enabling empirical analysis of their evolution and impact on development outcomes.

\end{longversion}

\section{Hypotheses}

The decomposition of prompts into three distinct but interrelated components in the prompt triangle allows us to understand \emph{what} developers communicate through prompts when working with chat-based AI coding assistants. These components capture requirement-oriented and solution-oriented information at different levels of abstraction. 
To also understand \emph{how} they use this content in the development process, we draw on requirements engineering theory. Following established SE practice~\cite{nuseibeh2000requirements}, we distinguish between three core activities: \emph{validation} (ensuring we are building the right thing by clarifying and confirming requirements), \emph{verification} (ensuring we are building the thing right by checking implementation correctness), and \emph{solution generation} (producing and exploring implementation approaches). These activities represent different purposes for which prompt content may be used during AI-assisted development. While the Prompt Triangle captures the semantic content of prompts, RE activities capture the pragmatic function of that content in the development workflow. This distinction is critical to understand how prompt composition influences code quality.

Based on this framework and prior empirical observations of iterative refinement in AI-assisted coding~\cite{barke2023grounded,chakraborty2026prompts}, we formulate the following hypotheses:
\begin{description}

\item[H1 (Prompt Evolution)]
During multi-turn AI-assisted coding sessions, the proportion of \textbf{Specific Solutions} content in prompts increases significantly across iterations relative to \textbf{General Solutions} and \textbf{Functionality and Quality}.

\item[H2 (User-driven Prompt Strategy)]
The distribution of Prompt Triangle components is significantly associated with developer characteristics, particularly programming experience and domain familiarity, with  programming experience and domain familiarity positively associated with \textbf{Specific Solution} proportion.

\item[H3 (RE Activity Alignment)]
Code quality is predicted by the alignment between prompt content and its RE purpose, with higher quality resulting from prompts where \textbf{Functionality and Quality} content is used for validation activities and \textbf{Specific Solution} content is used for verification activities.


\item[H4 (Temporal Alignment)]
Conversations that exhibit both the temporal prompt evolution pattern (H1) and RE activity sequence (H3) will achieve higher code quality than conversations with equivalent overall alignment but inconsistent temporal ordering.


\end{description}
H1 serves as descriptive validation that sets up H4, which is our key hypothesis about whether temporal alignment predicts success. While H3 tests whether alignment between prompt components and RE activities has an impact at all, H4 tests whether the point in time when alignment occurs matters.

\section{Study Overview}

We employ a \emph{multi-method data collection strategy} to construct a comprehensive corpus of developer–AI interactions. Our approach combines three complementary sources of data: (i) controlled experiments, (ii) community-sourced prompt exports, and (iii) mining of publicly available datasets.

Our triangulated design balances internal and ecological validity through three complementary data sources. Following Colnet et al.~\cite{colnet2024causal}, who review methods for combining randomised trials and observational studies, our controlled experiment ($n=30$) provides unconfounded causal estimates with full quality measures, while community uploads (target $n \geq 150$) and mined public datasets assess generalisability through observational data. Confirmatory hypothesis testing (H1--H4) is based on controlled experiment data; community uploads serve as co-primary replication data for H2 and supplementary validation for H1 and H3; mined datasets serve an explicitly exploratory role for pattern characterisation only.

\begin{longversion}
\begin{table}[tb]
\centering
\caption{Data Source Contribution to Hypothesis Testing}
\label{tab:data-sources}
\begin{tabular}{@{}P{2.4cm}P{3cm}P{3.9cm}P{3.4cm}@{}}
\toprule
\textbf{Hypothesis} & \textbf{Controlled Exp.} & \textbf{Community Uploads} & \textbf{Mined Data} \\
\midrule
\textbf{H1:} Prompt Evolution & Primary confirmatory test & Supplementary validation with sufficient turn length & Large-scale characterisation \\
\textbf{H2:} User Strategy & Co-primary test & Co-primary test (target combined n$\geq$150) & Not used (no demographics) \\
\textbf{H3:} RE Activity Alignment & Primary test + coding validation via think-aloud & Manual coding (20\% sample) + self-reported outcomes & Automated coding + proxy outcomes \\
\textbf{H4:} Temporal Pattern & Primary test with full quality measures & Pattern identification (quality via self-report/submission) & Pattern characterisation (no quality ground truth) \\
\bottomrule
\end{tabular}
\end{table}
\end{longversion}

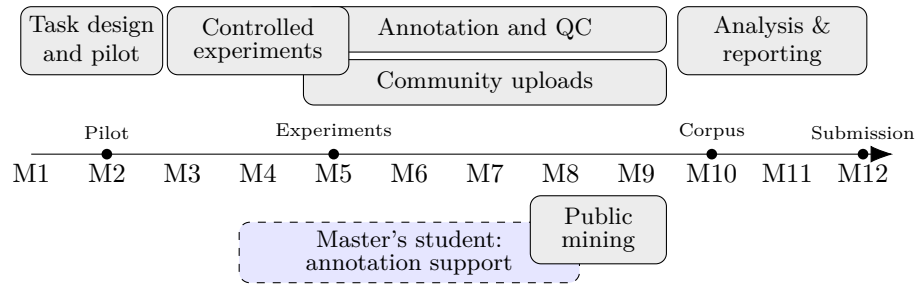
\begin{figure*}[t]
\centering
\begin{tikzpicture}[
x=1.0cm,
y=1cm,
phase/.style={draw, rounded corners, minimum height=0.9cm, inner sep=3pt, align=center, font=\small},
student/.style={draw, dashed, rounded corners, minimum height=0.8cm, inner sep=3pt, align=center, font=\small},
milestone/.style={circle, fill, inner sep=1.5pt}
]

\draw[-{Latex[length=3mm]}] (0,-0.2) -- (11.4,-0.2);

\foreach \x/\m in {0/M1,1/M2,2/M3,3/M4,4/M5,5/M6,6/M7,7/M8,8/M9,9/M10,10/M11,11/M12}
{
    \node[below] at (\x,-0.2) {\m};
}


\node[phase, fill=gray!15, minimum width=1.6cm] 
at (0.8,1.3) {\shortstack{Task design\\and pilot}};

\node[phase, fill=gray!15, minimum width=4.8cm, minimum height=0.6cm] 
at (6.0,0.75) {\shortstack{Community uploads}};

\node[phase, fill=gray!15, minimum width=4.8cm, minimum height=0.6cm] 
at (6.0,1.45) {\shortstack{Annotation and QC}};

\node[phase, fill=gray!15, minimum width=2.4cm] 
at (3.0,1.3) {\shortstack{Controlled\\experiments}};

\node[phase, fill=gray!15, minimum width=2.5cm] 
at (9.8,1.3) {\shortstack{Analysis \&\\reporting}};

\node[student, fill=blue!10, minimum width=4.5cm] 
at (5.0,-1.5) {\shortstack{Master's student:\\annotation support}};

\node[phase, fill=gray!15, minimum width=1.8cm] 
at (7.5,-1.2) {\shortstack{Public\\mining}};


\fill[milestone] (1.0,-0.2) circle (2pt);
\node[above=2pt, font=\scriptsize] at (1.0,-0.2) {Pilot};

\fill[milestone] (4.0,-0.2) circle (2pt);
\node[above=2pt, font=\scriptsize] at (4.0,-0.2) {Experiments};

\fill[milestone] (9.0,-0.2) circle (2pt);
\node[above=2pt, font=\scriptsize] at (9.0,-0.2) {Corpus};

\fill[milestone] (11.0,-0.2) circle (2pt);
\node[above=2pt, font=\scriptsize] at (11.0,-0.2) {Submission};

\end{tikzpicture}
\caption{Planned project timeline, starting M1=May'26 and submission by M12=April'27.}
\label{fig:timeline}
\end{figure*}

Figure~\ref{fig:timeline} summarizes the planned timeline of the study. The project begins with task design and pilot testing, followed by controlled experiments, community-sourced data collection, and public mining. Annotation and quality control partially overlap with data collection. A Master's student will contribute primarily to annotation support and corpus preparation.

\section{Data Collection}

Our methodology extends the approach pioneered by the DevGPT dataset~\cite{xiao2024devgpt}, which collected developer-ChatGPT conversations from publicly shared links. However, DevGPT primarily captures web-based ChatGPT interactions rather than conversations occurring within integrated development environments (IDEs). Since IDE-based AI assistants (e.g., GitHub Copilot, Cursor, JetBrains AI Assistant) represent increasingly common development workflows, our corpus focuses on these contextualised interactions.
Table~\ref{tab:collection-methods} summarizes the three collection approaches and their respective contributions to the corpus.

\begin{table*}[tb]
\centering
\caption{Comparison of data collection methods}
\label{tab:collection-methods}
\begin{tabular}{@{}P{2.1cm}P{2cm}P{5.2cm}P{3.4cm}@{}}
\toprule
\textbf{Method} & \textbf{Data Type} & \textbf{Strengths} & \textbf{Limitations} \\
\midrule
Controlled Experiments & Structured, curated & Complete metadata, ground truth requirements, comparable tasks & Limited scale, potential observer effects \\
Community Uploads & Real-world, diverse & Authentic usage, varied contexts, scalable & Incomplete metadata, self-selection bias \\
Public Mining & Opportunistic & No recruitment overhead, naturally shared & Sparse, variable quality \\
\bottomrule
\end{tabular}
\end{table*}

\subsection{Controlled Experiments}

\textbf{Participant Recruitment}
We will recruit $N=30$ software developers through university mailing lists, professional networks, and online communities. Participants must have $\geq$6 months programming experience and regular IDE-based AI assistant use. We aim for balanced representation across experience levels, collecting years of experience as a continuous variable alongside AI affinity, tool familiarity, and programming language proficiency. These variables are used continuously in regression analyses; categorical labels (junior/mid/senior) serve only as recruitment strata. For the repeated-measures ANOVA testing H3 (3 conditions) and H4 (4 conditions), power analysis using G*Power ($f = 0.35$, $\alpha = 0.05$, $power = 0.80$, correlation among repeated measures $= 0.5$) indicates minimum $N = 21$ for H3 and $N = 24$ for H4. Our target $N = 30$ provides adequate power for all planned within-subjects comparisons and accommodates potential dropout.

\textbf{Experimental Design}
Participants complete four 30-minute coding tasks: (1) \textbf{Feature Implementation (with AI)}: Add a feature to existing code using AI assistant and provided user stories; (2) \textbf{Code Refactoring (with AI)}: Improve code structure to meet quality requirements using AI; (3) \textbf{Algorithm Implementation (with AI)}: Implement specified algorithm from natural language description; (4) \textbf{Bug Fix (without AI)}: Diagnose and fix bug without AI assistance, establishing baseline RE activity rates. Tasks are selected to cover distinct SE activities that elicit different prompt compositions (requirements-heavy vs.\ solution-heavy) and varying AI interaction patterns. They are drawn from an open-source project with included requirement specifications, piloted for appropriate difficulty and 30-minute feasibility. Tasks are counterbalanced using Latin square design. Participants receive 15 minutes initial preparation to understand the codebase, with 5-minute breaks between tasks to mitigate fatigue effects over the 2:30h session.

\textbf{Data Collection Protocol}
Participants access a GitHub repository with codebase, task descriptions, and submission templates. We provide README setup instructions, dependency management files, optional containerised services (docker-compose), and environment verification scripts. Participants use their preferred IDE locally, acknowledging environmental heterogeneity (Section~\ref{sec:threats-to-validity}).
Chat histories are exported via native tool functionality (JSON/conversation export). For tools lacking export, we provide copy-paste templates. Sessions are conducted remotely via Zoom/Teams with screen sharing and recording for: (1) process observation, (2) think-aloud protocol capture, and (3) backup data if export fails (anticipated $<10\%$ cases). For the control task, participants verbalise their thought process, which we transcribe and code for RE activities comparable to AI conversations. Post-task questionnaires capture strategy and satisfaction data.
We collect per task: (a) exported chat history; (b) final code; (c) completion time; (d) screen recording; (e) think-aloud transcription; (f) questionnaire responses; (g) AI tool and version used.

\textbf{Annotation}
Two researchers independently code each conversation turn for: (1) \textbf{Prompt Triangle}: percentage of text per component (F\&Q, General Solutions, Specific Solutions; sum = 100\%); (2) requirement clarity (explicit/implicit/ambiguous); (3) turn type (query/refinement/clarification/confirmation/exploration); (4) code presence; (5) solution evolution stage (initial/debugging/refinement/finalization). Think-aloud verbalisations are coded for RE activities: validation (questions, assumptions, confirmations), verification (testing, requirement checking), and refinement (interpretation adjustments).

Inter-annotator agreement target: Cohen's $\kappa > 0.75$, indicating excellent agreement~\cite{landis1977measurement}. Disagreements resolved through discussion after initial alignment on a coded subset. A detailed coding manual following established content analysis methodology~\cite{saldana2021coding} with initial category definitions, decision rules for ambiguous cases, and worked examples will be developed during the pilot phase and preregistered before main data collection. The requirement clarity classification (explicit/implicit/ambiguous) is grounded in requirements quality attributes~\cite{boehm1984verifying} and requirements smell research~\cite{femmer2017rapid}; solution evolution stages (initial/debugging/refinement/finalization) follow iterative development models~\cite{basili1975iterative} and observed AI interaction patterns~\cite{barke2023grounded}.

\subsection{Community-Sourced Export and Upload}

We intend to develop a web-based platform enabling developers to voluntarily export and upload their IDE-based AI conversation histories. The platform includes:
\begin{itemize}
    \item Informed consent documentation compliant with IRB requirements
    \item Tool-specific export instructions for popular IDE assistants
    \item Drag-and-drop upload interface accepting multiple file formats
    \item Automated anonymisation preview allowing participants to review and redact sensitive information
    \item Demographic and usage context survey to collect programming, AI coding, and task experience
\end{itemize}
\textbf{Recruitment and Incentivisation}
Recruitment will proceed in waves over six months via academic networks, professional developer communities (Reddit, LinkedIn, Discord), and conference presentations. We will provide early access to research findings and optional acknowledgment in dataset documentation. We acknowledge heterogeneity in tools, tasks, and contexts in community uploads; inclusion criteria (complete demographics, $\geq 3$ turns, $\geq 100$ words) and the unified processing pipeline ensure comparability for analysis.

\textbf{Sample Size and Power Analysis}
Our target sample of $n \geq 150$ community uploads supports co-primary testing of H2 alongside controlled experiment data. Power analysis for the planned mixed-effects hierarchical regression using G*Power (approximating the mixed-effects structure as hierarchical regression) indicates that a combined sample of $n \geq 120$ (30 controlled + 90 community minimum) provides 80\% power ($\alpha = 0.05$) to detect a medium effect size ($f^2 = 0.10$) for the incremental $\Delta R^2$ contribution of developer characteristics when testing 4--5 predictors. The target of $n \geq 150$ accounts for approximately 20\% anticipated exclusion due to incomplete demographics, insufficient conversation length, or failed quality control. Should community uploads fall below $n = 90$ after filtering, we will proceed with reduced power but supplement interpretation with effect size confidence intervals and Bayesian sensitivity analyses.

\noindent\textbf{Data Processing Pipeline}
Each uploaded conversation passes through a five-stage pipeline:
\begin{enumerate}
    \item \textbf{Format normalisation}: Conversations in various formats (JSON, CSV, plain text, screenshots with OCR) are converted to a unified structure preserving temporal order and speaker identification.
    
    \item \textbf{Automated quality checks}: Minimum length verification (3+ turns), code presence detection, and programming language identification. Conversations failing any criterion are flagged for manual review.
    
    \item \textbf{PII detection}: Named entity recognition (spaCy) combined with pattern matching for emails, API keys, and proprietary identifiers. Flagged content is manually reviewed; participant privacy takes precedence over data retention.
    
    \item \textbf{Manual review}: A stratified random sample of 15\% of uploads (stratified by data source, programming language, and conversation length) undergoes manual quality verification by two independent researchers, with inter-rater agreement calculated using Cohen's kappa.
    
    \item \textbf{Semi-automated annotation}: Fine-tuned transformer models predict Prompt Triangle component proportions per turn. Predictions with confidence below 0.7 are queued for manual annotation.
\end{enumerate}

\subsection{Public Mining}

Public mining provides opportunistic access to naturally shared developer--AI interactions without recruitment overhead, capturing developers who might not volunteer for formal studies.

\noindent\textbf{Source Identification}
To complement recruited data, we intend to systematically mine publicly shared IDE-based AI conversations from:
\begin{itemize}
    \item GitHub issues and discussions mentioning specific AI tools
    \item Developer blogs and tutorial platforms (Medium, Dev.to)
    \item Video content (YouTube coding tutorials with AI assistants)
    \item Social media platforms (Twitter/X, Reddit, LinkedIn)
    \item Technical forum discussions (Stack Overflow, specialised Discord servers)
\end{itemize}
Given the exploratory role of public mining, sample size will be determined by data availability (target $\geq 200$ conversations).

\textbf{Collection Methodology}
We will employ keyword-based searches (e.g., ``GitHub Copilot conversation,'' ``Cursor AI help,'' ``AI assistant code'') combined with manual verification. Searches will be conducted bi-weekly over the data collection period. Only content explicitly shared publicly by developers will be included.

A preliminary search confirms available sources (e.g., the DevGPT dataset~\cite{xiao2024devgpt} and GitHub Copilot conversations shared in issue trackers), though public conversations are not always complete and may reference outdated tool versions. We will record publication dates to allow temporal filtering and treat incomplete conversations as valid for turn-level analysis. We acknowledge that publicly shared conversations may be biased toward noteworthy or extreme examples; this is mitigated by the explicitly exploratory role of mined data and by triangulating any observed patterns against controlled experiment results.

\textbf{Ethical Considerations}
For mined content, we verify that data was intentionally shared publicly, check platform licensing and terms of use, and apply additional anonymisation to protect original authors. Where feasible, we attempt to contact original posters for explicit consent and honour all removal requests.

\begin{longversion}
\subsection{Unified Data Schema and Integration}

All collected conversations, regardless of source, were transformed into a unified JSON schema to facilitate analysis. The schema captures:

\begin{lstlisting}[language=json, caption=Core data schema structure, label=lst:schema, breaklines=true]
{
  "conversation_id": "unique-identifier",
  "source": "controlled|uploaded|mined",
  "metadata": {
    "collection_date": "ISO-8601-timestamp",
    "ide_tool": "tool-name",
    "programming_language": ["list"],
    "developer_experience": "level|unknown",
    "task_type": "task-category|unknown",
    "has_ground_truth_requirement": boolean
  },
  "conversation": [
    {
      "turn_number": integer,
      "speaker": "developer|ai",
      "content": "text-and-code",
      "timestamp": "if-available",
      "annotations": {
        "requirement_clarity": "explicit|implicit|ambiguous|null",
        "contains_code": boolean,
        "turn_type": "type|null",
        "solution_stage": "stage|null"
      }
    }
  ],
  "outcome": {
    "final_code": "if-available",
    "task_completion": "success|failure|partial|unknown"
  }
}
\end{lstlisting}

This schema allows for varying levels of metadata richness across sources while maintaining structural consistency.
\end{longversion}

\subsection{Quality Control}

Quality control combines automated filtering and manual verification:%
\begin{itemize}
    \item \textbf{Length filter}: Minimum 5 conversation turns, concrete number will be adjusted based on a sensitivity analysis
    \item \textbf{Language detection}: Conversations in supported programming languages (Python, JavaScript, Java, C++, C\#, Go, Rust, TypeScript)
    \item \textbf{Code presence}: At least one turn containing code snippets
    \item \textbf{Duplicate detection}: Identify and remove duplicate conversations using fuzzy matching
    \item \textbf{Completeness scoring}: Flag conversations with missing turns or corrupted content
\end{itemize}
\textbf{Manual Verification}
A stratified random sample of 15\% of all uploaded and mined conversations will undergo manual review by two researchers to verify quality and appropriateness. Conversations failing manual review criteria will be excluded from the final corpus. The automated filters will be extended if patterns appear during this manual review to exclude other data points with similar issues.

\begin{longversion}
\subsection{Resulting Corpus Characteristics}

The final corpus comprises [N] conversations totaling [M] individual turns and [K] code snippets. Table~\ref{tab:corpus-stats} presents detailed statistics by collection method.

\begin{table*}[t]
\centering
\caption{Corpus statistics by collection method}
\label{tab:corpus-stats}
\begin{tabular}{lrrrr}
\toprule
\textbf{Source} & \textbf{Conversations} & \textbf{Turns} & \textbf{Code Snippets} & \textbf{Fully Annotated} \\
\midrule
Controlled & 150 & 2,847 & 1,203 & 100\% \\
Uploaded & 3,421 & 48,392 & 19,847 & 25\% \\
Mined & 287 & 3,118 & 1,456 & 10\% \\
\midrule
\textbf{Total} & \textbf{3,858} & \textbf{54,357} & \textbf{22,506} & \textbf{31\%} \\
\bottomrule
\end{tabular}
\end{table*}

The corpus spans [X] programming languages, with Python (38\%), JavaScript (24\%), and Java (15\%) being most prevalent. Task types include feature implementation (32\%), debugging (28\%), code understanding (18\%), testing (12\%), and refactoring (10\%).
\end{longversion}

\section{Data Analysis}
\label{sec:data-analysis}

\textbf{Analysis Plan for H1 (Prompt Evolution)}
Since prompt component proportions are compositional (summing to 1), we employ Dirichlet mixed-effects regression~\cite{maier2014dirichletreg}, the appropriate model family for such data~\cite{aitchison1986compositional}. The model is:
\begin{equation*}
(\text{F\&Q}_{it},\, \text{GS}_{it},\, \text{SS}_{it}) \sim \text{Dirichlet}(\boldsymbol{\alpha}_{it}), \quad \log(\alpha_{k,it}) = \beta_{0k} + \beta_{1k} \cdot \text{Turn}_t + u_{ik}
\end{equation*}
where $i$ indexes participants, $t$ indexes turns, $k \in \{\text{F\&Q}, \text{GS}, \text{SS}\}$, and $u_{ik}$ is a participant-level random intercept. H1 predicts $\beta_{1,\text{SS}} > 0$ (Specific Solutions increases) and $\beta_{1,\text{F\&Q}} < 0$ (F\&Q decreases), with non-overlapping 95\% CIs. We test coefficients using likelihood ratio tests. Sensitivity analyses include: (1) linear mixed-effects models on each proportion separately for comparison with prior work, and (2) repeated-measures ANOVA for participants completing all turns. We report coefficients, 95\% CIs, and pseudo-$R^2$.

\textbf{Analysis Plan for H2: User-Driven Prompt Strategy}
We conduct hierarchical regression to test how developer characteristics predict prompt composition, controlling for task type. Using controlled experiment data ($n=30$), the multivariate regression model is:
\begin{equation*}
\text{SS\_proportion}_{ij} = \beta_0 + \boldsymbol{\gamma} \cdot \mathbf{Controls}_j + \boldsymbol{\beta} \cdot \mathbf{DevChar}_i + u_i + \varepsilon_{ij}
\end{equation*}
where $\mathbf{Controls}_j$ (Step~1) includes task type, complexity, and domain; $\mathbf{DevChar}_i$ (Step~2) includes programming experience, domain familiarity, and AI tool experience; and $u_i$ is a participant random intercept. The dependent variable is the Specific Solutions proportion. H2 predicts Step~2 $\Delta R^2 \geq .10$ ($p < .05$), with positive associations between experience/familiarity and Specific Solutions proportion. Since the dependent variable is a bounded proportion, we conduct sensitivity analysis using beta mixed-effects regression to verify robustness to distributional assumptions.

For between-subjects comparison, mixed-effects regression combines controlled experiment and community uploads (target $n \geq 120$; inclusion criteria: complete demographics, $\geq 3$ turns, $\geq 100$ words). Fixed effects include experience, familiarity, task type, and data source; random effect is participant nested in source. If community $n < 120$, we proceed with reduced power. Supplementary Spearman correlations describe characteristic-component relationships. We test multivariate normality (Mardia's test) and multicollinearity (VIF $< 5$), applying transformations if violated. Sensitivity analyses examine experience as continuous vs. categorical and explore non-linear relationships.

\textbf{Analysis Plan for H3 (RE Activity Alignment)}
We employ dual-coding where each prompt segment receives both a Prompt Triangle component code and an RE activity code (validation, verification, solution generation). Two coders independently annotate all controlled experiment conversations (participant--AI and think-aloud verbalizations), targeting Cohen's $\kappa > 0.75$ per dimension. We develop automated pre-labelling to scale coding to community and mined data; manual control tasks are coded for RE activities only.

We operationalise \emph{code quality} as a composite of (1)~functional correctness, measured by pass rate of predefined test cases for each task, and (2)~structural quality, assessed via static analysis metrics (maintainability index, cyclomatic complexity, code smells) from an open-source tool (e.g., SonarQube Community Edition). The composite score is the standardised mean of these two dimensions.

For each conversation, we calculate alignment scores: (1) F\&Q-Validation alignment and (2) SS-Verification alignment. The primary regression model is:
\begin{equation*}
\text{Quality}_{ij} = \beta_0 + \beta_1 \cdot \text{FQ\_Val}_{ij} + \beta_2 \cdot \text{SS\_Ver}_{ij} + \boldsymbol{\gamma} \cdot \mathbf{X}_{ij} + u_i + \varepsilon_{ij}
\end{equation*}
where $\mathbf{X}_{ij}$ includes overall component proportions, conversation length, task type, and developer experience, and $u_i$ is a participant random effect. Hypothesis confirmation requires at least one significant positive alignment coefficient ($p < .05$).

Repeated-measures ANOVA compares three within-subjects conditions from the controlled experiment ($n=30$), classified post-hoc by median-splitting participants' alignment scores within the AI-assisted tasks: (1)~Manual baseline (the non-AI control task), (2)~Low-alignment AI (below-median alignment), (3)~High-alignment AI (above-median alignment). We predict Manual $\approx$ Low-alignment $<$ High-alignment, with high-alignment significantly exceeding both (Cohen's $d \geq 0.5$). Community uploads use dual-coding with self-reported quality; mined data use proxy measures with automated coding validated on 10\% manual samples. Sensitivity analyses test robustness across sources and explore synergistic effects.

\textbf{Analysis Plan for H4 (Temporal Pattern)}
To test whether temporal alignment predicts quality beyond overall alignment, we employ hierarchical regression dividing conversations into early/late turns (50\% split), corresponding to the two-stage model where early turns focus on specification (F\&Q-dominant) and late turns shift to refinement (SS-dominant). Step~1 controls for overall alignment, task type, and experience ($\boldsymbol{\gamma} \cdot \mathbf{Step1}_{ij}$). Step~2 adds: Stage~1 alignment (F\&Q with solution generation early), Stage~2 alignment (SS with validation/verification late), and temporal pattern score $\text{TP}_{ij} = \sqrt{\text{Stage1}_{ij} \times \text{Stage2}_{ij}}$:
\begin{equation*}
\text{Quality}_{ij} = \beta_0 + \boldsymbol{\gamma} \cdot \mathbf{Step1}_{ij} + \beta_{\text{S1}} \cdot \text{Stage1}_{ij} + \beta_{\text{S2}} \cdot \text{Stage2}_{ij} + \beta_{\text{TP}} \cdot \text{TP}_{ij} + u_i + \varepsilon_{ij}
\end{equation*}
Support requires $\Delta R^2 \geq .05$ ($p < .05$) and positive $\beta_{\text{TP}}$. Sensitivity analysis replaces the binary split with continuous turn-weighted alignment. Repeated-measures ANOVA ($n=30$) compares four post-hoc conditions (Manual, Low/Low, High/Low, High/High AI), predicting Manual $\approx$ Low/Low $<$ High/Low ($d \geq 0.4$) $<$ High/High ($d \geq 0.3$). Community and mined data provide exploratory pattern prevalence.

\section{Threats to Validity}
\label{sec:threats-to-validity}

We organise threats to validity following the framework by Wohlin et al.~\cite{wohlin2012experimentation}, addressing construct validity, internal validity, external validity, and reliability as shown in 
Table \ref{tab:threats}.

\begin{table}[tb]
\small
\caption{Summary of Threats to Validity and Mitigation Strategies}
\label{tab:threats}
\begin{tabular}{@{}p{1.8cm}p{4.2cm}p{7cm}@{}}
\toprule
\textbf{Category} & \textbf{Threat} & \textbf{Mitigation} \\
\midrule
\multirow{2}{2cm}{Construct} 
& Prompt Triangle subjectivity & Detailed codebook, $\kappa > 0.75$, third annotator for disagreements \\
& Code quality measurement limits & Multiple metrics (alignment, static analysis, corrections); acknowledge dimensions not captured \\
\midrule
\multirow{3}{2cm}{Internal} 
& Observer/Hawthorne effects & Emphasise studying AI not participants; 15-min familiarisation; observational data sources \\
& Task selection bias & Tasks based on literature prevalence; multi-source data captures diverse real-world tasks \\
& Learning effects & Counterbalanced task order; task order as covariate in analyses \\
\midrule
\multirow{3}{2cm}{External} 
& Sample representativeness & Report demographics; compare to industry stats; diverse data sources \\
& Task authenticity & Real open-source codebase; varied requirement formats; time constraints acknowledged \\
& Temporal/tool evolution & Document tool versions; framework designed to be tool-agnostic \\
\midrule
\multirow{2}{2cm}{Reliability} 
& Annotation consistency & Calibration every 50 conversations; detect drift by comparing annotation halves \\
& Tool heterogeneity & Comprehensive format normalisation; explicit quality control (Section 5.4) \\
\bottomrule
\end{tabular}
\end{table}

\begin{longversion}
We organise threats to validity following the framework by Wohlin et al.~\cite{wohlin2012experimentation}, addressing construct validity, internal validity, external validity, and reliability.

\subsection{Construct Validity}

Prompt Triangle Operationalisation: The decomposition of prompts into Functionality and Quality, General Solutions, and Specific Solutions relies on manual annotation by trained coders. Despite establishing inter-rater reliability (target $\kappa$ > 0.75), subjective interpretation of prompt intent may introduce inconsistencies. To mitigate this, we will develop a detailed codebook with examples and conduct iterative calibration sessions. Additionally, we will employ a third annotator to resolve persistent disagreements.

Code Quality Measurement: Our assessment of generated code quality combines multiple metrics (requirement-implementation alignment, static analysis, post-generation corrections). However, these metrics may not fully capture code maintainability, performance, or security. We acknowledge that quality is multidimensional and that our selected metrics represent a pragmatic subset. Future work should incorporate additional quality dimensions such as test coverage and runtime performance.

\subsection{Internal Validity}

Observer Effects (Hawthorne Effect): Participants in controlled experiments may alter their natural prompting behavior due to awareness of being observed and recorded. To minimise this threat, we emphasise that we are studying the AI interaction rather than evaluating participant performance. We also allow a 15-minute familiarisation period to reduce initial self-consciousness. The inclusion of community-sourced and mined data provides observational comparison data less susceptible to this threat.

Task Selection Bias: The three experimental tasks (feature implementation, refactoring, algorithm implementation) may not represent the full spectrum of AI-assisted development activities. We selected these tasks based on their prevalence in prior literature and their ability to elicit different prompting strategies. However, we acknowledge that other task types (debugging, documentation, test generation) may yield different patterns. The multi-source data collection partially addresses this by capturing diverse real-world tasks.

Learning Effects: Participants complete three tasks, potentially improving their prompting strategies across tasks due to learning. We employ counterbalancing of task order across participants to distribute learning effects evenly. Additionally, we will analyse whether improvement patterns differ from our hypothesised evolution patterns by including task order as a covariate in our analyses.

AI Model Variability: Participants may use different AI assistants (GitHub Copilot, Cursor, Claude Code) with varying capabilities, which could confound the relationship between prompt structure and code quality. We will record the specific AI tool and version used and include this as a control variable in our statistical analyses. Sensitivity analyses will examine whether relationships hold across different AI platforms.

\subsection{External Validity}

Participant Representativeness: Our sample of N=30 developers may not represent the broader population of software developers. In particular, developers who volunteer for research studies may be more experienced with AI tools or more motivated than typical users. We will report detailed demographic characteristics and compare them to industry statistics where available. The community-sourced and mined data help address this by capturing more diverse developer populations.

Task Authenticity: Controlled experimental tasks, despite being based on realistic scenarios, occur in an artificial setting with time constraints (30 minutes per task) and predefined requirements. Real-world development involves ambiguous requirements, organizational constraints, and longer timeframes. To enhance ecological validity, we deliberately use varied requirement formats (user stories, quality requirements, natural language descriptions) and select tasks from real open-source projects. The complementary real-world data sources provide additional ecological validity.

Temporal Validity: AI coding assistants evolve rapidly, and prompting patterns observed in this study may become obsolete as models improve and new interaction paradigms emerge. We will explicitly document the versions of AI tools used and acknowledge that findings represent a snapshot of current practices. The conceptual framework (Prompt Triangle) is designed to be tool-agnostic and may remain relevant even as specific tools evolve.

Programming Language and Domain Specificity: Our experimental tasks and mined data focus on a limited set of programming languages (Python, JavaScript, Java, C++, C\#, Go, Rust, TypeScript). Prompting patterns may differ substantially in other languages or specialised domains (e.g., embedded systems, scientific computing). We will analyse whether patterns differ across the included languages and discuss generalisability limitations in our interpretation.

\subsection{Reliability}

Annotation Consistency: Manual annotation of prompts is vulnerable to coder drift over time, where annotators gradually change their interpretation of coding schemes. To address this, we will conduct regular calibration sessions every 50 conversations and recalibrate the codebook if $\kappa$ falls below $0.75$. We will also analyse the first and second halves of annotations separately to detect drift.

Data Processing Reproducibility: The multi-stage data processing pipeline (format normalisation, PII detection, annotation) involves both automated and manual steps that may be difficult to replicate. We will document all processing steps in detail, version-control our processing scripts, and release them alongside the dataset. For automated components (PII detection, quality filtering), we will report precision and recall metrics.

Platform and Tool Heterogeneity: Allowing participants to use their preferred IDE and AI assistant increases external validity but reduces experimental control. Different export formats and capabilities across platforms may lead to inconsistent data quality. We mitigate this through comprehensive format normalisation and explicit quality control procedures (Section 4.4), though some information loss is unavoidable.

Replication Data Limitations: While we plan to release a three-tier dataset, the most sensitive data (Tier 3) will only be available through collaboration, limiting independent replication. This trade-off balances open science principles with privacy and ethical obligations. We will ensure that all statistical analyses can be validated using Tier 1 and Tier 2 data, with Tier 3 serving only for exploratory extensions.
\end{longversion}

\section{Ethical Considerations}
We intend to submit the revised research methodology based on this document and the feedback provided by the ESEM Registered Reports Track reviewers to the University of Bayreuth Institutional Review Board. All participants in controlled experiments and community uploads will have to provide informed consent. Key ethical measures include voluntary participation with right to withdraw, comprehensive data anonymisation removing names, email addresses, company identifiers, and proprietary code patterns, secure data storage with encryption on institutional servers, participant option to review contributed data before finalisation, compliance with GDPR and relevant data protection regulations, and transparent data usage and retention policies.

\pagebreak
\bibliographystyle{plainurl}
\bibliography{prompt-triangle-methodology}

\end{document}